# Physical realization of topological Roman surface by spin-induced ferroelectric polarization in cubic lattice

Guangxiu Liu[1,2,†], Maocai Pi[3,4,†], Long Zhou[1], Zhehong Liu[1,2], Xudong Shen[5,1], Xubin Ye[1,2], Shijun Qin[1,2], Xinrun Mi[3,4], Xue Chen[3,4], Lin Zhao[1,2], Bowen Zhou[1,2], Jia Guo[1,2], Xiaohui Yu[1,2], Yisheng Chai[3,4,*], Hongming Weng[1,2,5,*], Youwen Long[1,2,5,*]

[1]*Beijing National Laboratory for Condensed Matter Physics, Institute of Physics, Chinese Academy of Sciences, Beijing 100190, China*

[2]*School of Physics, University of Chinese Academy of Sciences, Beijing 100049, China*

[3]*Center of Quantum Materials and Devices, Chongqing University, Chongqing 401331, China.*

[4]*Low Temperature Physics Laboratory and Chongqing Key Laboratory of Soft Condensed Matter Physics and Smart Materials, College of Physics, Chongqing University, Chongqing 401331, China*

[5]*Songshan Lake Materials Laboratory, Dongguan, Guangdong 523808, China*

*Corresponding email: yschai@cqu.edu.cn, hmweng@iphy.ac.cn, ywlong@iphy.ac.cn

[†]These authors contributed equally to this work

Topology, a mathematical concept in geometry, has become an ideal theoretical tool for describing topological states[1-3] and phase transitions[4]. Many topological concepts have found their physical entities in real or reciprocal spaces identified by topological/geometrical invariants[2,5,6], which are usually defined on orientable surfaces such as torus and sphere. It is natural to quest whether it is possible to find the physical realization of more intriguing non-orientable surfaces. Herein, we show that the set of spin-induced ferroelectric polarizations in cubic perovskite oxides $A\mathrm{Mn_3Cr_4O_{12}}$ ($A$ = La and Tb) resides on the topological Roman surface[7], a non-orientable two-dimensional manifold formed by sewing a Möbius strip edge to that of a disc. The induced polarization may travel in a loop along the non-orientable Möbius strip or orientable disc depending on how the spin evolves as controlled by external magnetic field. Experimentally, the periodicity of polarization can be the same or the twice of the rotating magnetic field, being well consistent with the orientability of disc and Möbius strip, respectively. This path dependent topological magnetoelectric effect presents a way to detect the global geometry of the surface and deepens our understanding of topology in both mathematics and physics.

Topological physics is one of the most successful combinations of mathematics and physics. As one of the branches of pure mathematics, topology has recently been employed to describe and understand many intriguing physical phenomena such as the quantum Hall effect,[1,8] topological insulators, topological semimetals, and topological superconductors.[9-11] These topological electronic states can be classified by various topological invariants, such as Chern number $N$,[2,5] topological class $Z_2$,[12] $Z_4$,[13] and Euler class $\chi$.[14,15] It is well known that Chern number is defined as the integral of vector field—Berry curvature over the first Brillouin zone of a two-dimensional (2D) periodic system, which is looked as a torus ($\boldsymbol{T^2}$) and serves as the base manifold. On the other hand, the magnetic spins and electric dipoles in real space can form topological objects or defects such as skyrmions,[16,17] polar skyrmions,[18] and polar merons.[19] These can be identified with an integer winding number[6] based on the field of spin or electrical dipole orientations over a base manifold of sphere ($\boldsymbol{S^2}$). In this sense, topological invariants are the global characteristics and the topology-related physical properties are typically robust against perturbations.[9,10,12]

Besides torus and sphere, there are other three fundamental 2D manifolds found in algebraic topology as listed in Supplementary Table 1, namely Möbius strip, Klein bottle, and Roman surface. Torus and sphere are orientable and have two sides. Berry curvature flux enclosed by them can be nonzero quantized value and taken as topological invariant like Chern number. In contrast, the other three are non-orientable surfaces and have just one side. The flux of any vector field through them is zero. The well-known example is that the one-side Klein bottle cannot be filled with water, but a two-side bottle can. In addition, the non-orientable surfaces are featured with a geometrical property that a person can be mirror-reflected when he travels along a special loop path on the surface, such as along the Möbius strip ($\boldsymbol{M}$) without crossing the boundary, and he recovers after a second loop travel. It is intriguing to realize a physical system that can evolve globally on a non-orientable surface to reflect such unique geometrical property.

In the real space, a Möbius strip with a single boundary has been visualized in nanocrystal/DNA bands.[20,21] In reciprocal space of electronic[22-24] and acoustic crystals,[25,26] Möbius insulator has been proposed and realized. Under proper nonsymmorphic symmetry or projective translation symmetry, the Bloch wavefunctions of its boundary states have $4\pi$ periodicity instead of $2\pi$, which mimics a Möbius strip. Recently, the nonorientability has been regarded as $Z_2$ gauge charge to describe the topological defect of shear deformation on a Möbius strip.[27] However, how to evolve a physical quantity smoothly and globally on a non-orientable surface and to detect its intrinsic geometry is still challenge. Furthermore, a non-orientable 2D surface without a boundary must self-intersect in three-dimensional (3D) space, as shown by the Klein bottle[28] ($K$, one-sided bottle) and Roman surface[7] ($RP^2$, real projective plane). This inhibits the physical realization and exhibition of topological behaviors in 3D real space. Therefore, it is very natural to ask whether Roman surface, the simplest non-orientable 2D manifold without boundary, can be realized or not by some physical entity and even can demonstrate some exotic topological properties or phenomena?

The Roman surface was named by Jakob Steiner, who discovered it during his visit to Rome in early 1844. It is composed of a Möbius strip sewing to the edge of a disc and has quite high tetrahedral $T_d$ symmetry, being different from other non-orientable 2D surfaces. There are three self-intersecting double lines $D_i$ ($i = a$, $b$, and $c$ are three mutually perpendicular Cartesian axes) that form a right trihedron in a Euclidean 3D phase space. Each double line ends with two pinch points and they all intersect at the triple point (see Fig. 1 and Supplementary Fig. 1). The triple point at the Roman surface indicates the intersection of three surfaces. The analytical equation of the Roman surface $\boldsymbol{RP^2}$ in mathematics is

$$x^2y^2 + x^2z^2 + y^2z^2 - \delta xyz = 0, \qquad (1)$$

where $\delta$ is a constant and $(x, y, z)$ is a point in $a$, $b$, $c$ coordinate system. As a 2D manifold, there are three types of fundamental closed loops on a Roman surface, and other closed loops can be formed by their combinations. As shown in Fig. 1c, the first type of loop is the red one that passes through at least one point twice (e.g., this point can be the triple point). It winds twice to close the path on this non-orientable surface since it rides on a Möbius strip, a component of Roman surface. The other two types of loop are the blue and green ones. They are topologically different from the red one since they do not pass any point twice and wind once to close the path. The blue loop can only wind one of the double lines $D_i$ ($i = a, b, c$) by $W_i = 1$ times. The green loop does not enclose any double line and it is topologically trivial. The physical entity that matches Eq. 1 may provide an opportunity to realize Roman surface and exhibit path dependent properties. Therefore, realizing Roman surface is quite intriguing and might extend the territory in the field of topological physics.

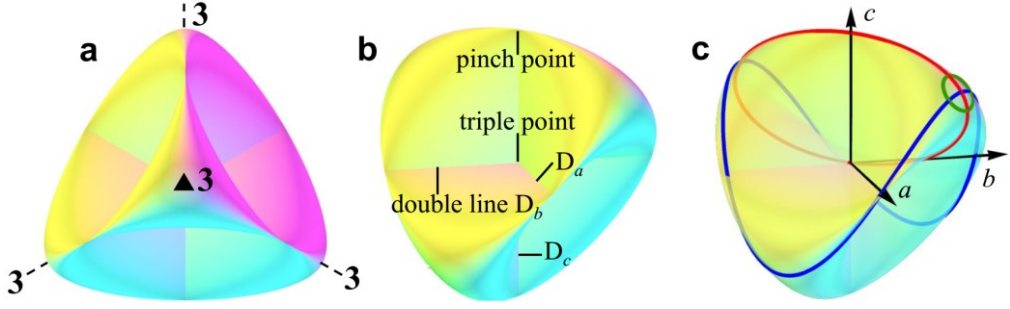

**Fig. 1 | Schematic diagram of Roman surface. a**, **b**, Roman surface with four three-fold axes (**3**), one triple point, six pinch points, and three double lines ($D_a$, $D_b$, $D_c$). **c**, Three geometrically different fundamental loops on a Roman surface. The red loop effectively winds twice to pass the two sides of the triple point since it rides on a Möbius strip being part of the Roman surface. The blue one winds the double line $D_c$ once to have $W_c = 1$. The green loop does not enclose any double line with zero $W_i$ ($i = a, b, c$). Both blue and green loops only wind once to close the path.

Magnetoelectric (ME) multiferroics with spin-induced ferroelectric polarization ($P$) have received much attention owing to the coupled spin and electrical dipole orders,[29] where the electric polarization can be switched using a magnetic field ($B$), and conversely the magnetization ($M$) can be switched using an electric field ($E$),[29,30] giving rise to significant ME effects. A few physical mechanisms have been proposed for ME multiferroics,[31-34] providing a fertile and unique playground for investigating the topological properties as a mapping between the spin vector space and electric dipole vector space. Although some structurally and ferroelectrically coupled vortex domain patterns have been observed in hexagonal $R$MnO$_3$ ($R$ = rare earth or Y, In) and BiFeO$_3$,[29,35] spin-dependent topological multiferroics remain to be elucidated, as do topological ME effects with versatile controls of $P$ by $B$ or $M$ by $E$. Here, we discovered that the set of spin-induced ferroelectric polarizations in $A$-site ordered cubic perovskites $A$Mn$_3$Cr$_4$O$_{12}$ can be described by the non-orientable Roman surface, as deduced through a 3D symmetry analysis. Various angle-dependent $P$ oscillations under different magnetic fields were observed and can be understood based on the changes in the local topological structure and the symmetry of the magnetoelectricity.

Figure 2a shows the powder x-ray diffraction (XRD) pattern and the refinement result of the newly synthesized TbMn$_3$Cr$_4$O$_{12}$ (TMCO) polycrystalline sample at room temperature. The Rietveld analysis demonstrates that the compound crystallizes in an *A*-site ordered perovskite structure with the chemical formula $AA'_3B_4O_{12}$ (see Fig. 2b).[36,37] The space group is a cubic *Im*-3 with central symmetry, where Tb and Mn are 1:3 ordered at special Wyckoff sites 2*a* (0, 0, 0) and 6*b* (0, 0.5, 0.5), and Cr and O at special 8*c* (0.25, 0.25, 0.25) and 24*g* (*x*, *y*, 0) sites, respectively. Compared with a simple $ABO_3$ perovskite, the introduction of a transition metal into the $A'$ site in the ordered $AA'_3B_4O_{12}$ perovskite causes the formation of square-coordinated $A'O_4$ units and heavily tilting $BO_6$ octahedra, as shown in Fig. 2b and described in detail elsewhere.[38,39] The refined structural parameters of TMCO are listed in Supplementary Table 2. According to the refined Cr-O and Mn-O bond lengths, the bond valence sum calculations illustrate that the valence states of Mn and Cr are both close to +3, in agreement with the reported isostructural compound $LaMn_3Cr_4O_{12}$ (LMCO).[40] Moreover, with decreasing temperature to 5 K, there is no trace of the structural phase transition in TMCO (see Supplementary Fig. 2).

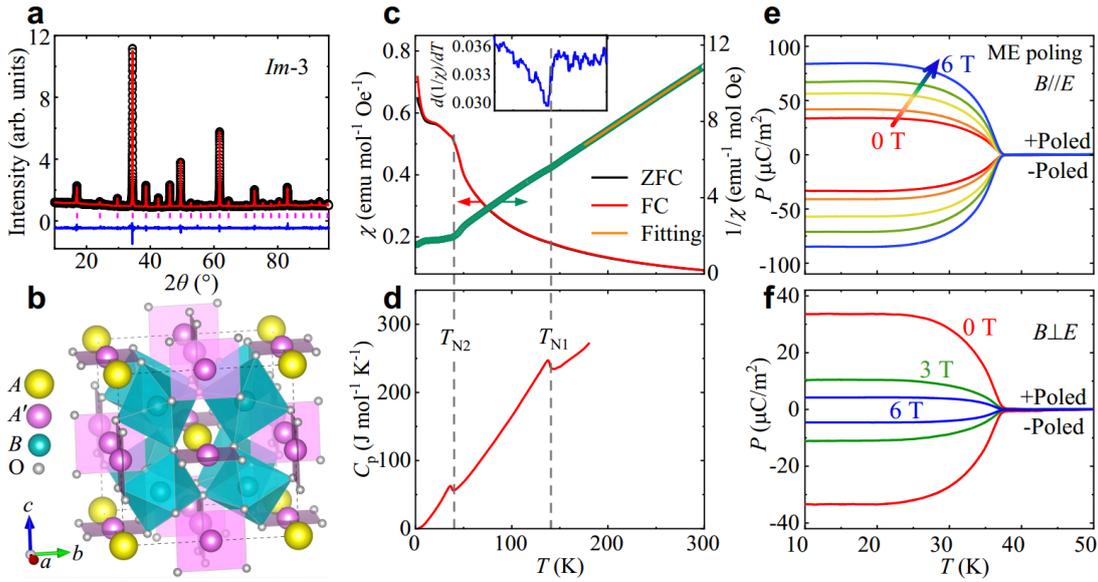

**Fig. 2 | Structure characterization, magnetic, specific heat and polarization measurements for TMCO. a**, XRD pattern measured at room temperature and the Rietveld refinement results. The observed (black circles), calculated (red line), and difference (blue line) values are shown. The ticks indicate the allowed Bragg reflections with space group *Im*-3. **b**, Crystal structure of *A*-site ordered perovskite $AA'_3B_4O_{12}$ with *Im*-3 symmetry. **c**, Temperature dependence of magnetic susceptibility $\chi$ measured at 0.1 T and its inverse $\chi^{-1}$. Both zero-field-cooling (ZFC) and field-cooling (FC) modes were adopted to measure $\chi$. The yellow line shows the Curie-Weiss fitting above 175 K. The fitted effective moment $\mu_{eff}$ (15.52 $\mu_B$ f.u.$^{-1}$) is comparable to the theoretical one (15.04 $\mu_B$ f.u.$^{-1}$) by considering the contribution from Tb$^{3+}$, Mn$^{3+}$ and Cr$^{3+}$ ions. The inset shows the derivative of inverse susceptibility versus temperature. **d**, Temperature dependence of specific heat $C_p$ measured at zero field. **e, f**, Temperature dependent $P$ after poling the sample under an electric field (= $E_{pole}$) and different magnetic fields $B$, where $B$ was not removed until the temperature increased to 50 K. For $B//E$ and $B \perp E$ configurations, both +Poled and -Poled conditions were measured.

Figure 2c and d show the temperature dependence of the magnetic susceptibility and specific heat of TMCO, respectively. With decreasing temperature to $T_{N1} \approx 136$ K, a sharp $\lambda$-type anomaly is observed in the specific heat, suggesting the occurrence of a second-order long-range magnetic phase transition. The derivative of the inverse susceptibility also displays a variation at $T_{N1}$ (see the inset of Fig. 2c). Upon further cooling to $T_{N2} \approx 36$ K, both susceptibility and specific heat show a clear anomaly, illustrating a second long-range magnetic transition. Below $T_{N2}$, the specific heat does not show any visible anomaly, indicating the absence of spin ordering of $Tb^{3+}$ at temperatures down to 2 K. The inverse susceptibility above 175 K follows the Curie-Weiss law (Fig. 2c). The fitted Weiss temperature of −28.0 K is indicative of the AFM interactions. The linear isothermal magnetization behaviors within ±2 T further confirm the AFM orders occurring at $T_{N1}$ and $T_{N2}$, as shown in Supplementary Fig. 3a. However, at higher fields, a metamagnetic transition occurs near 2.5 T at temperatures below $T_{N1}$ $\approx$ 136 K, as demonstrated by the field derivative of magnetization, indicating field-induced variation of spin alignment (Supplementary Fig. 3b). Similar to LMCO,[40] at zero field, the $B$-site $Cr^{3+}$ spins are antiferromagnetically ordered in a collinear $G$-type fashion at $T_{N1}$, and the $A'$-site $Mn^{3+}$ sublattice experiences another $G$-type AFM ordering at $T_{N2}$ in the current TMCO (see Fig. 3a for a detailed spin structure). The total spin structure composed of $Cr^{3+}$ and $Mn^{3+}$ magnetic sublattices with spin moments along the [111] direction forms a type-II polar magnetic point group of **31′**, which breaks the space inversion symmetry. Therefore, a spin-induced ferroelectric phase transition is expected to occur at $T_{N2}$.

Corresponding to the AFM phase transition, a frequency-independent dielectric peak is observed at $T_{N2} \approx$ 36 K in TMCO, whereas no dielectric anomaly emerges at $T_{N1}$ (Supplementary Fig. 4a and b). When a magnetic field $B$ and an electric field $E$ are simultaneously applied to measure the dielectric constant, an apparent anisotropic behavior can be observed. Specifically, a sharp dielectric peak is always observed with $B$ up to 9 T, if $B//E$ (Supplementary Fig. 4c). By contrast, $B$ significantly suppresses the dielectric peak in the $B \perp E$ configuration, whereas the dielectric anomaly nearly disappears at 6 T (Supplementary Fig. 4d). These results strongly suggest the spin-induced ferroelectric polarization as well as the anisotropic ME properties in TMCO. Furthermore, when the pyroelectric current $I_p$ was measured to obtain the temperature-dependent polarization, a sharp ferroelectric phase transition was found to occur at $T_{N2}$ (Fig. 2e), as expected from the formation of the polar magnetic point group mentioned above. Moreover, the polarization is completely reversible without changing the magnitude if the sign of the poling electric field ($E_{pole}$) is reversed, further confirming the spin-induced ferroelectric phase transition in TMCO. Similar to the dielectric measurements, applying a magnetic field with a $B//E$ configuration can significantly enhance the polarization from 33.3 μC/m² at 0 T to 84.4 μC/m² at 6 T (Fig. 2e). In the $B \perp E$ configuration, however, the spin-induced polarization is suppressed to 10.7 μC/m² at $B = 3$ T, and only 4.51 μC/m² at $B = 6$ T (Fig. 2f). All experimental results indicate that TMCO has the same crystal and magnetic structures and spin-induced polarization as those observed in LMCO.[40]

Based on the crystal and spin structure symmetry analysis, we now show that the peculiar ME multiferroicity of $A$Mn$_3$Cr$_4$O$_{12}$ can be well described through the topological Roman surface **$RP^2$**. As mentioned above, both TMCO and LMCO have a cubic crystal symmetry of $Im$-3, where inversion centers are located on each Cr atom, with four 3-fold rotation axes along four body diagonals, and three mirror planes perpendicular to the three Cartesian axes (Fig. 3a). The total magnetic structure composed of Cr$^{3+}$ spins (we use Néel vector $S_{Cr}$ to trace the coherent motion of spins in AFM ordering of Cr ions) and Mn$^{3+}$ spins (Néel vector $S_{Mn}$) forms a polar magnetic point group **$31'$** with $S_{Mn}//S_{Cr}//$[111], as shown in Fig. 3a. According to the proposed anisotropic symmetric exchange mechanism[41] (see Supplementary Section 1) originating from the spin-orbit coupling, such special ME coupling induces polarization $\vec{P}$ along the [111] direction. To describe a more general relationship between the spin and polarization directions, we assign $S_{Cr} = (\mu_{Cr}, \theta, \varphi)$ and $S_{Mn} = (\mu_{Mn}, \theta, \varphi)$. Both are on a sphere **$S^2$** (Fig. 3a and c), where $\theta \in [0, \pi]$ and $\varphi \in [-\pi, \pi]$ with fixed magnitudes $\mu_{Cr}$ and $\mu_{Mn}$. Thus, the polarization $P$ is expressed as a function of spins $\vec{P}(S_{Mn}, S_{Cr}) = \vec{P}(\theta, \varphi)$ mapping from $S^2$ to $RP^2$. Based on the strict symmetrical analysis,[42] as described in the Supplementary Sections 2 and 3, one can see that $\vec{P}(\theta, \varphi)$ has the following form:

$$\vec{P} = \beta\mu_{Cr}\mu_{Mn}\Omega[\sin(2\theta)\sin(\varphi), \sin(2\theta)\cos(\varphi), \sin^2\theta\sin(2\varphi)], \qquad (2)$$

where the coefficients $\beta$ and $\Omega$ depend on the temperature and crystal properties. The polarization vector $\vec{P}(\theta, \varphi) = (P_a, P_b, P_c)$ can be rewritten using an analytical equation as follows:

$$P_a^2 P_b^2 + P_a^2 P_c^2 + P_b^2 P_c^2 - 2\beta\mu_{Cr}\mu_{Mn}\Omega P_a P_b P_c = 0. \qquad (3)$$

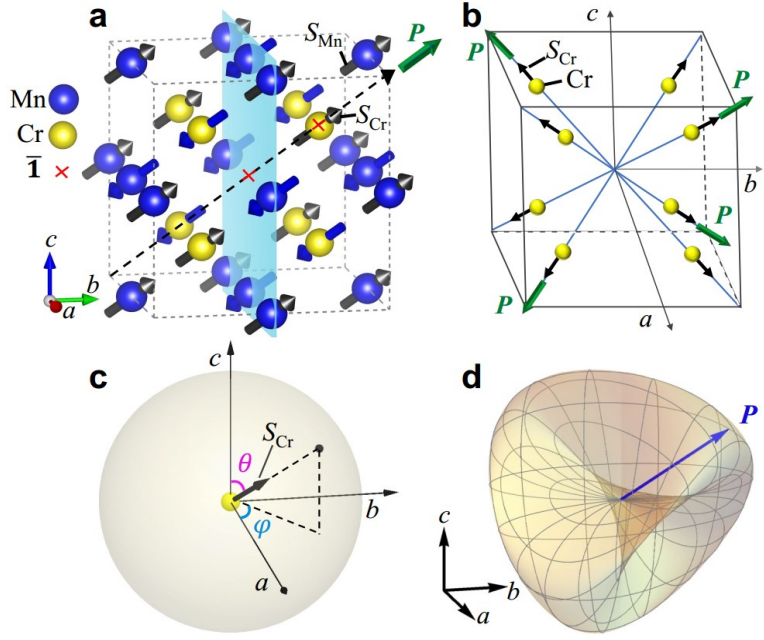

**Fig. 3 | Schematic diagram of crystal, magnetic, and ferroelectric structures in $A$Mn$_3$Cr$_4$O$_{12}$.**
**a**, Crystal and spin configurations where Cr and Mn atoms are presented in yellow and blue, respectively, and O atoms are omitted for clarity. Two space inversions $\bar{1}$ located at the center and at the Cr atom, one of the mirror planes (see the light-blue plane), and one of the 3-fold rotation axes (see the dash line) are also indicated. All spins composed of Mn and Cr magnetic sublattices are parallel or antiparallel to the [111] direction. AFM sublattices of Cr and Mn are denoted by Néel vectors $S_{Cr}$ and $S_{Mn}$, respectively, and the induced polarization $P$ is shown by the arrow in green. **b**, Distribution of $P$ vectors along four of the eight diagonal directions. The yellow sphere and the black arrow denote the Cr atom and its spin moment. $S_{Mn}$ is always parallel to $S_{Cr}$ and it is not shown for clarity. Thus, AFM configuration composed of Mn and Cr sublattices is only indicated by $S_{Cr}$. The black cube represents the crystal unit cell. **c**, Orientation of $S_{Cr}$ represented in spherical coordinates ($\mu_{Cr}$, $\theta$, $\varphi$). **d**, The complete set of allowed spin-induced $P$ vectors in three dimensions by the magnetic structure shown in **a** is a Roman surface.

Clearly, Eq. 3 is a specific case of Eq. 1 with $x = P_a, y = P_b, z = P_c$, and $\delta = 2\beta\mu_{Cr}\mu_{Mn}\Omega$. This result indicates that the induced $P$ vectors in $A$Mn$_3$Cr$_4$O$_{12}$ may constitute a non-orientable Roman surface **RP²**, as shown in Fig. 3d. In this cubic multiferroic family, the collinear $G$-type AFM locks Néel vectors $S_{Mn}$ and $S_{Cr}$ parallelly and they pass through all eight equivalent diagonal directions coherently (see Supplementary Section 3 for details). Only four of the eight diagonal directions are allowed for the polarization $P$ vectors, which form a regular tetrahedron and agrees well with the tetrahedral symmetry of the Roman surface (Fig. 3b). It is noted that there is another AFM configuration with $S_{Mn}$ antiparallel to $S_{Cr}$ as shown in Supplementary Fig. 5, which leads to a set of polarization vectors $P$ forming the Roman surface with negative $\delta$ and having similar features as the positive one discussed here.

The spin-polarization relationship in the cubic perovskites $A$Mn$_3$Cr$_4$O$_{12}$ leads to topological multiferroicity owing to the topological and symmetry properties of the Roman surface discussed above. The mapping from **S²** to **RP²** covers the Roman surface twice because $S_{Cr}$ and $-S_{Cr}$ induce the identical $P$ vectors. This is very similar to the spin-momentum locking effect on the Fermi surface enclosing Weyl nodes. The Hamiltonian around Weyl node maps the momentum space (Fermi sphere **S²**) to spin space (another sphere named as Bloch sphere). If the enclosed Weyl node is of topological charge one or two, the momentum moves around the node once and the spin rotates once or twice, respectively. The relative rotation directions of momentum and spin are determined by the sign of the charge.[43] In this sense, a closed evolution loop of Néel vectors $S_{Cr}$ on the sphere may map to a non-trivial path of the $P$ vector on the Roman surface showing topological ME effects.

The red loop in Fig. 1c passes through the triple point of the Roman surface twice where $P = 0$. This can be induced by the evolution of $S_{Cr}$ on sphere $S^2$ that includes the six special points where sphere intersects $a$-, $b$-, $c$-axis (shown in Fig. 4a). For a longitudinal great circle of the sphere with $\varphi$ fixed as shown in Fig. 4a, $S_{Cr}$ passes through the special points twice (the north and south poles) if $\varphi$ is not 0, $\pi/2$, $\pi$ nor $3\pi/2$. This leads to a double-winding ellipse trajectory of the $P$ vector passing through the triple point twice (see Fig. 4b). The induced $P$ rotates twice while $S_{Cr}$ travels one loop. When $\varphi = 0$, $\pi/2$, $\pi$ or $3\pi/2$, the trajectories of $P$ collapse to the double lines and passes through the special points four times (see Fig. 4b). One can see that only the magnitude of $P$ vibrates, and the period is twice of loop evolution of $S_{Cr}$.

When the Néel vectors evolve along the latitudinal parallels with polar axis being $c$-axis and fixed $\theta$ value ($\theta$ is not 0 or $\pi/2$) as shown in Fig. 4c, the induced saddle-like trajectory of $P$ on Roman surface only winds around the $c$-axis (Fig. 4d). This trajectory of $P$ is nontrivial in the sense that the $W_c = 1$. Its projection onto the $ab$-plane is a circle. The rotation direction, clockwise or counter- clockwise, of the projection of $P$ vector on the circle is opposite to that of the Néel vector. It is noted that the rotation period of $P$ vector is the same as $S_{\text{Cr}}$ instead of double of that in cases of Fig. 4b. Similarly, when $S_{\text{Cr}}$ winds either the $a$- or $b$-axis once, the corresponding $P$ vector loop on Roman surface is characterized by $W_a = 1$ or $W_b = 1$, respectively, as indicated in Fig. 4e and f.

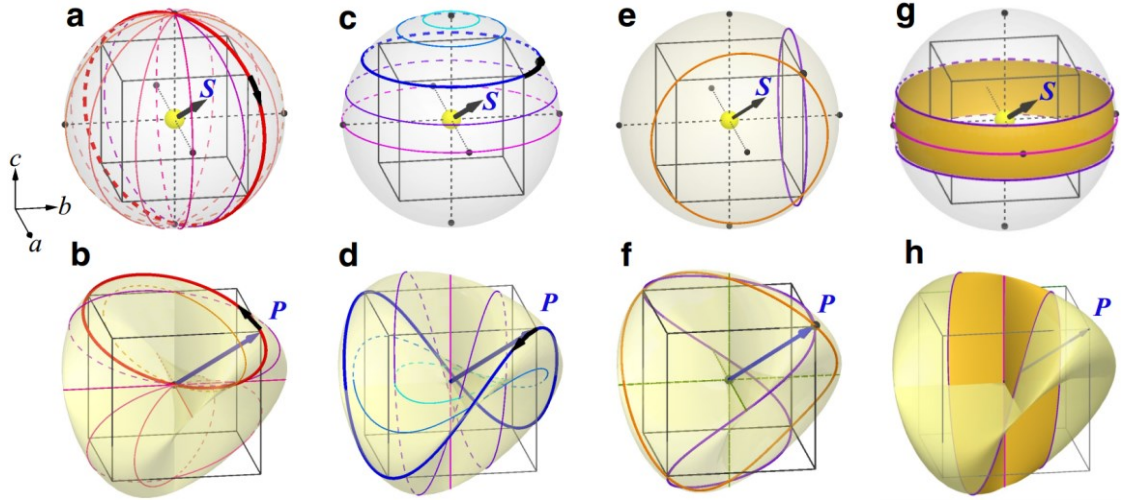

**Fig. 4 | Various loops on sphere of Néel vectors (top panels) and the induced fundamental polarization trajectories on Roman surface (bottom panels). a**, Longitudinal (fixed $\varphi$) path of Néel vector rotation mode with a series of selected $\varphi$ values and continuously varying $\theta \in [0, 2\pi]$. **b**, Trajectories of $P$ induced by loops in **a**. **c**, Latitudinal (fixed $\theta$) Néel vector rotation mode with a series of selected $\theta$ values and continuously varying $\varphi \in [-\pi, \pi]$. **d**, Trajectories of $P$ induced by loops in **c**. The six special points, the intersections between sphere and $a$-, $b$- and $c$-axis, are illustrated in black point in **a** and **c**. The thicker lines in **a** and **c** represent two special cases where four diagonal directions are passed through; the resultant trajectories of $P$ are indicated by the thicker lines in **b** and **d**, respectively. Rotation directions of $S_{\text{Cr}}$ and $P$ are indicated by black arrows toward opposite directions away from [111] direction. **e**, Two closed loops winding the $a$ and $b$ axes once are shown in orange and purple, respectively. **f**, Two closed trajectories of $P$ on the Roman surface induced by the loops in **e**. The loops with $W_a = 1$ (orange) and $W_b = 1$ (purple). **g**, A band included the equator of the Néel vector sphere is mapped to **h**, a self-crossed Möbius strip as a part of the Roman surface. The residual upper and lower caps are mapped to a disc on the Roman surface at the same time. All the loops in **b**, **d** and **f** are fundamental loops defined in Fig. 1c.

It is notable that nonzero $W_i$ corresponds to the component of $P$ vector perpendicular to $i$-axis reverses $W_i$ times during one loop evolution. The green circle shown in Fig. 1c is topologically trivial with zero values of $W_i$. It can be mapped from a loop on $S_{Cr}$ sphere that satisfies these conditions: there is no any special point being wound and $-S_{Cr}$ will not be passed by for any $S_{Cr}$ on the path. To reveal the underlying topological origins of double-winding and single-winding paths in Figs. 4b, 4d and 1c, we find that a strip composed by the neighborhoods of equator and itself on the spin sphere (Fig. 4g) is mapped to a Möbius strip (Fig. 4h) on the Roman surface. In the meantime, the residual upper and lower caps on sphere are both mapped to a saddle-like disc. Thus, according to the orientable and non-orientable nature of the disc and the Möbius strip, the paths on them will wind once and twice to form a closed loop, respectively.

We experimentally present the topological ME effects in TMCO and LMCO polycrystalline samples, which possess identical crystal and spin structures. First, the samples were cooled down to 10 K under $E_{pole}//y$ at zero $B$. Subsequently, we removed the electric field $E_{pole}$ and measured the $B$-dependent $P$ by sweeping $B$ in either the $B//y$ or $B//x$ configuration (Fig. 5a). In both configurations, quadratic ME effects were dominant in TMCO (Fig. 5b). This is consistent with the **31′** point group with independent time reversal symmetry where a quadratic rather than a linear ME effect is allowed. To investigate the postulated topological ME effects in TMCO, a rotating $B$ was applied to tune the orientation of the Néel vector $S_{Cr}$ continuously. The $B$ direction dependence of $P_y$ under selected field strengths by changing $\psi$ (initially $\psi = 0$ with $B//y$; Fig. 5c–f) was measured after electrical poling down to 10 K. The $\psi$ dependence of $P_y$ exhibited double-angle sinusoidal modulation curves at $B$ within ~2 T without changing the sign (Fig. 5c). This is a direct manifestation of the intrinsic quadratic ME behaviors in this system. Based on the Monte Carlo simulations, we find that, under a relatively small rotating $B$ without inducing metamagnetic transition, the Néel vector $S_{Cr}$ will slightly tilt away from the diagonal direction, leading to a topologically trivial path on the sphere, but it is noted that $S_{Cr}$ itself is double period of the rotating $B$ (see Supplementary Section 4 and Supplementary Fig. 6a). The resulting loop of $P$ vector on the Roman surface is also a small perturbation to its original position with zero values of $W_i$ (Supplementary Fig. 6b), as exemplified by the green loop in Fig. 1c. Therefore, the $S_{Cr}$ will travel along green-type path twice when the $B$ rotates by 360° (Supplementary Fig. 7). This results in a double winding of the loop on Roman surface, consistent with the experimentally observed double-angle sinusoidal modulation of $P_y$ for small $B$, indicating the trivial mapping from $S_{Cr}$ to $P$ along the green loop.

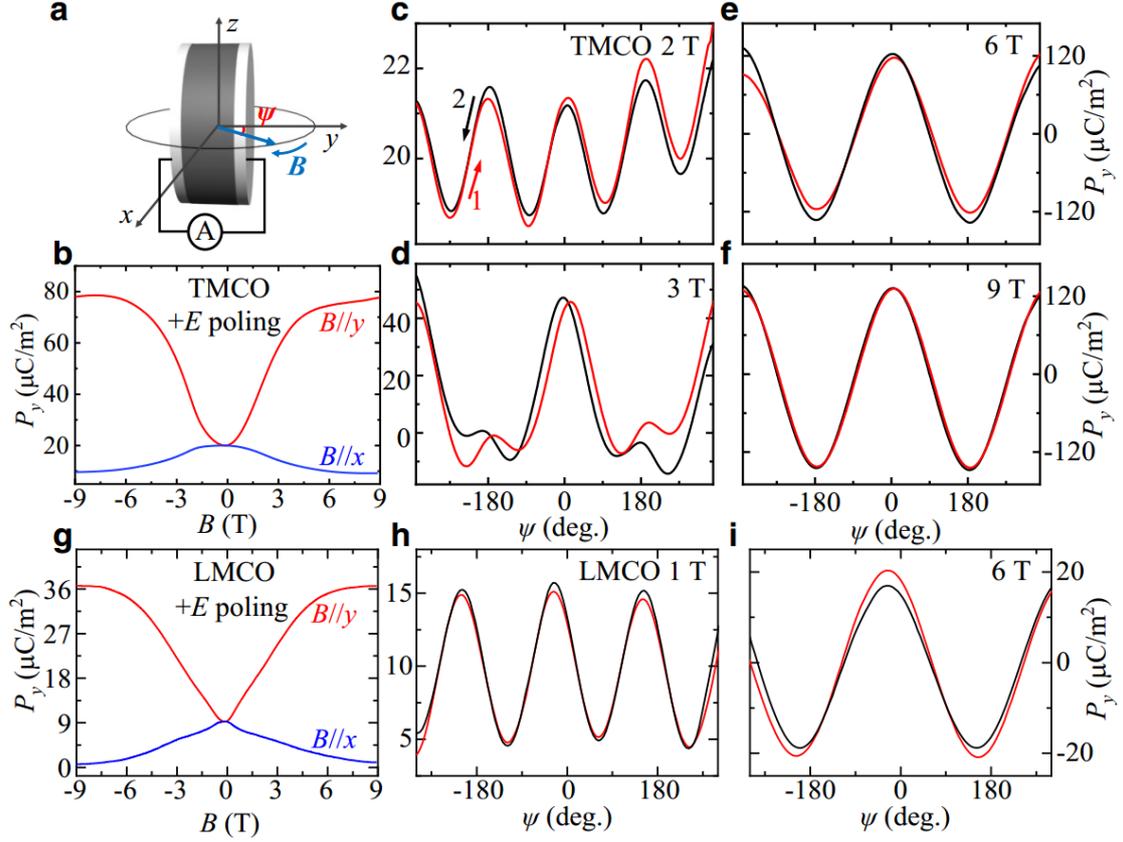

**Fig. 5 | Polarization variations under sweeping and rotating $B$ at 10 K for TMCO and LMCO. a**, Schematic illustration of measurements for ME current under rotating $B$ within $xy$-plane; $\psi$ denotes the angle between $B$ and the $y$-direction. **b**, **g**, Magnetic-field-dependent $P_y$ at $T = 10$ K after $+E$ poling for $B//E$ and $B \perp E$ configurations of TMCO and LMCO, respectively. $P_y$ is defined as the component of $P$ along the $y$-axis. **c-f**, $\psi$-dependent polarization of TMCO at $B = 2$ T (**c**), 3 T (**d**), 6 T (**e**), and 9 T (**f**). The red and black arrows indicate angle an increase and decrease, respectively. **h**, **i**, $\psi$-dependent polarization of LMCO at $B = 1$ T (**h**) and 6 T (**i**). All polarizations were integrated from the ME current as a function of time.

For $B$ values greater than 3 T, a sinusoidal $P_y$ variation appeared, owing to the metamagnetic transition occurring at a critical field of approximately 2.5 T (Supplementary Fig. 3b). At $B$ = 6 and 9 T, the $P_y$ value can be fully reversed with a pure sinusoidal behavior (Fig. 5e and f). The energy barrier of magnetic anisotropy among diagonal directions can be overcome by increasing the Zeeman energy, and a spin flop transition will finally occur at a larger rotating $B$ (> 3 T) to induce a nontrivial loop of the Néel vector on the sphere. The loop of Néel vector winds its principal axis under a sufficiently high $B$, which induces $P$ vector loop with one of $W_i$ = 1 ($i = a, b, c$) on Roman surface accordingly. This can lead to reversing of the $P$ components perpendicular to $i$-axis like the blue loop in Fig. 1c. The pure sinusoidal $P$ behavior at 6 and 9 T in the TMCO can thus be understood.

As shown in Fig. 5d, both single- and double-angle sinusoidal $P_y$ features coexist at $B$ = 3 T with clear hysteresis behavior. The $P$ = 0 values have been reached multiple times in one period. There is a possibility that the loops of Néel vector pass some of the crystal axes, and some naught $P$ values occur during a 360° rotation of $B$ like the red loop in Fig. 1c. Note that, owing to the polycrystalline specimens used in experiments, it is impossible to determine the detailed grain configurations or $P$ rotating paths of each grain for the TMCO. In the isostructural LMCO, all of the above-mentioned ME behaviors can be reproduced (Fig. 5g-i), indicating the universality of these phenomena in the cubic multiferroic perovskite family of $A\text{Mn}_3\text{Cr}_4\text{O}_{12}$.

Moreover, we propose that the topological nontrivial rotation of $P$ can be realized in the multiferroic domain walls of $A\text{Mn}_3\text{Cr}_4\text{O}_{12}$ without external magnetic field. As shown in Supplementary Fig. 8, there are three types of multiferroic domains with $S_{\text{Mn}}$ parallel to $S_{\text{Cr}}$ on considering the Néel vectors prefer to eight diagonal directions, i.e. the 71, 109 and 180 deg. AFM domains induced 109, 109 and 360 deg. ferroelectric domains, respectively. In particular, the induced 360 deg. ferroelectric domain wall represents a half winding of Möbius strip since a 180 deg. rotation of $S_{\text{Cr}}$ leads to one loop of $P$ and mirror reflected on Roman surface (Supplementary Fig. 8c). This type of multiferroic domain wall cannot be destroyed by external electric field due to the fact that the polarizations in two neighboring ferroelectric domains can be coherently aligned to the electric field direction while the related AFM configurations will always be opposite. This would be a new type of topological object unique in $A\text{Mn}_3\text{Cr}_4\text{O}_{12}$ which deserves further investigation.

In summary, we investigated the unusual spin-polarization relationship in $A$-site-ordered cubic perovskite oxides $A\text{Mn}_3\text{Cr}_4\text{O}_{12}$. The uncommon polarization occurring in the cubic system is strongly coupled with the inversion symmetry breaking caused by a $G$-type collinear AFM ordering composed of $Mn^{3+}$ and $Cr^{3+}$ magnetic sublattices. We discovered that the magnetoelectric behavior, namely the mapping from spin to the induced polarization, can be looked as a physical entity realizing Roman surface, one of the non-orientable 2D manifolds. This is essentially different from other topological objects defined on orientable manifolds, such as Chern insulator defined on torus and skyrmion defined on sphere. The nonorientability of Roman surface leads to topological ME effects depending on the evolution of path. In our experiments, different kinds of angle-dependent polarization are observed in polycrystalline TMCO and LMCO under different magnetic fields. The evolution paths of spin vectors are controlled by magnetic field and the induced $P$ evolves along orientable or non-orientable surface to show different topological properties of them. This study provides an interesting fact that the topology has a close and deep relationship with the physical world, opening up a new avenue to achieve more topological objects from orientable manifold to non-orientable one.

for multiferroicity. *Phys. Rev. B* **93**, 174416 (2016).

**Methods**

Polycrystalline TMCO was synthesized via a solid-state reaction under high temperature and high pressure. $Tb_4O_7$, $MnO$, $Cr_2O_3$, and $Mn_2O_3$ powders were used as starting materials at a mole ratio of 1:2:8:5. The finely mixed reactants were packed into a gold capsule with 4.0 mm in long and 2.8 mm in diameter and then treated at 1300 K and 7 GPa for 30 min on a cubic-anvil-type high-pressure apparatus.

Powder XRD was performed using a Huber diffractometer with Cu K$\alpha$1 radiation at 40 kV and 30 mA. The diffraction data were collected in the 2$\theta$ angle range from 10° to 100°. A program from the general structure analysis system was used to refine the XRD data based on the Rietveld method.[44] Raman spectra were collected on a T64000 spectrometer using a liquid-nitrogen-cooled charge-coupled device. Volume Bragg gratings of 600G and 488 nm laser were applied to measure the spectra.

Magnetic susceptibility and magnetization were measured using a superconducting quantum interference device magnetometer (Quantum Design, MPMS-7T). Zero-field-cooling and field-cooling modes were adopted to measure the susceptibility at a magnetic field of 0.1 T. Magnetization was measured between -14 and 14 T at selected temperatures. Specific heat measurements were performed using a physical property measurement system (Quantum Design, PPMS-9T) at zero field.

A hard disk-shaped piece with about 2.0 mm in diameter and 0.2 mm in thickness was used to measure the relative dielectric constant on an Agilent-4980A LCR meter. Silver paste was used as electrodes. Subsequently, the sample was used to measure the pyroelectric current and isothermal polarized current under different magnetic fields on a Keithley 6517B electrometer. To measure the pyroelectric current for TMCO, the sample was poled with an electric field $E_{pole} = \pm 11.7$ kV/cm and a magnetic field from 50 to 10 K; subsequently, the electric field was removed. After waiting for 30 min in an electric short-circuit, the pyroelectric current data were collected at temperatures ranging from 10 to 50 K at a speed of 2 K/min. The polarization $P$ was obtained by integrating the current data as a function of time. For the measurements of isothermal polarized current, the sample was poled by an electric field of approximately 8.33 kV/cm for TMCO and 8.68 kV/cm for LMCO down to 10 K. After electrical poling procedure, a magnetic field was applied to the sample to measure the isothermal polarized current and then the magnetic field dependent polarization or the angle dependent polarization can be obtained by the integration of the current.

**Acknowledgements** This study was supported by the Beijing Natural Science Foundation (Grant No. Z200007), the National Key R&D Program of China (Grant No. 2018YFE0103200, 2018YFA0305700, 2016YFA0300600), the National Natural Science Foundation of China (Grant No. 11934017, 51772324, 11921004, 11904392, 11674384, 11925408), the Chinese Academy of Sciences (Grant No. XDB33000000, QYZDB–SSW–SLH013, GJHZ1773), Fundamental Research Funds for the Central Universities (2020CDJQY-A056, 2018CDJDWL0011), and Projects of President Foundation of Chongqing University (2019CDXZWL002). The authors acknowledge the fruitful discussion with X. H. Chen.

**Author contributions** Y.L., Y.C. and H.W. conceived and led this project. M.P., X.C., and Y.C. performed theoretical calculations. G.L., X.Ye, B.Z., and J.G. synthesized the samples. G.L., L.Zhou, X.S., and X.M. performed dielectric constant and polarization measurements and data analysis. Z.L., S.Q., L.Zhao, and X.Yu performed structural and magnetic measurements. G.L., M.P., H.W., Y.C., and Y.L. wrote the manuscript. All the authors discussed the results and commented on the manuscript.

# Supplementary Information for

**Physical realization of topological Roman surface by spin-induced ferroelectric polarization in cubic lattice**

Guangxiu Liu, Maocai Pi, Long Zhou, Zhehong Liu, Xudong Shen, Xubin Ye, Shijun Qin, Xinrun Mi, Xue Chen, Lin Zhao, Bowen Zhou, Jia Guo, Xiaohui Yu, Yisheng Chai[*], Hongming Weng[*], Youwen Long[*]

[*]Corresponding author. E-mail: yschai@cqu.edu.cn (Y.C.); hmweng@iphy.ac.cn (H.W.); ywlong@iphy.ac.cn (Y.L.)

Content

# 1. Microscopic origin of polarization in cubic $A$Mn$_3$Cr$_4$O$_{12}$

The most three well-known microscopic origin mechanisms of spin induced polarization ($P$) can be written as[1,2]:

$$P = P^{MS}(S_1 \cdot S_2)e_{12} + P^{SP}e_{12}\times(S_1\times S_2) + P^{ORB}[e_1(e_1 \cdot S_1)^2 - e_2(e_2 \cdot S_2)^2] \quad (S1)$$

where the spins $S_1$ and $S_2$ reside on the adjacent sites along the unit vector $e_1$, $e_2$. The $e_1$ and $e_2$ are unit vectors pointing from site $1$ and site $2$ to ligand atom, respectively. $P^{MS}$, $P^{SP}$ and $P^{ORB}$ are scalars determined by the exchange interaction, the Dzyaloshinskii-Moriya (DM) interaction and $p$-$d$ hybridization, respectively. All the above three mechanisms can be expressed in two spin tensor in matrix form as:

$$P_{exchange}^{\alpha\beta\gamma} = P^{MS}\begin{pmatrix} e_x,e_y,e_z & 0,0,0 & 0,0,0 \\ 0,0,0 & e_x,e_y,e_z & 0,0,0 \\ 0,0,0 & 0,0,0 & e_x,e_y,e_z \end{pmatrix} \quad (S2)$$

$$P_{DM}^{\alpha\beta\gamma} = P^{SP}\begin{pmatrix} 0,0,0 & e_y,-e_x,0 & e_z,0,-e_x \\ -e_y,e_x,0 & 0,0,0 & 0,e_z,-e_y \\ -e_z,0,e_x & 0,-e_z,e_y & 0,0,0 \end{pmatrix} \quad (S3)$$

$$P_{hybridization}^{\alpha\beta\gamma} = P^{ORB}\begin{pmatrix} e_xe_xe_x,e_xe_xe_y,e_xe_xe_z & e_xe_ye_x,e_xe_ye_y,e_xe_ye_z & e_xe_ze_x,e_xe_ze_y,e_xe_ze_z \\ e_ye_xe_x,e_ye_xe_y,e_ye_xe_z & e_ye_ye_x,e_ye_ye_y,e_ye_ye_z & e_ye_ze_x,e_ye_ze_y,e_ye_ze_z \\ e_ze_xe_x,e_ze_xe_y,e_ze_xe_z & e_ze_ye_x,e_ze_ye_y,e_ze_ye_z & e_ze_ze_x,e_ze_ze_y,e_ze_ze_z \end{pmatrix} \quad (S4)$$

From the polarization expressions of Eqs. S7 and S15, we can see that the effective components $P^{a,b,c}+P^{b,c,a}+P^{c,a,b}+P^{a,c,b}+P^{b,a,c}+P^{c,b,a}$ in $P_{ij}^{\alpha\beta\gamma}$ do not match with any known microscopic mechanism in Eqs. S2-S4. Therefore, the intrinsic polarization of TbMn$_3$Cr$_4$O$_{12}$ TMCO is governed by a new mechanism, anisotropic symmetric exchange mechanism due to spin orbital coupling[3].

## 2. Derive of the matrix form of two spin tensors

In general, the third-order local ME two-spin tensor $P_{12}^{\alpha\beta\gamma}$ of Cr-Mn bond $\langle 1, 2 \rangle$ in Supplementary Fig. 5a can be expressed as in the matrix form[4]:

$$P_{12}^{\alpha\beta\gamma} = \begin{pmatrix} \{P^{a,a,a}, P^{a,a,b}, P^{a,a,c}\} & \{P^{a,b,a}, P^{a,b,b}, P^{a,b,c}\} & \{P^{a,c,a}, P^{a,c,b}, P^{a,c,c}\} \\ \{P^{b,a,a}, P^{b,a,b}, P^{b,a,c}\} & \{P^{b,b,a}, P^{b,b,b}, P^{b,b,c}\} & \{P^{b,c,a}, P^{b,c,b}, P^{b,c,c}\} \\ \{P^{c,a,a}, P^{c,a,b}, P^{c,a,c}\} & \{P^{c,b,a}, P^{c,b,b}, P^{c,b,c}\} & \{P^{c,c,a}, P^{c,c,b}, P^{c,c,c}\} \end{pmatrix} \quad \text{(S5)}$$

After considering the five symmetry operations (3-fold along [1,1,1], space inversion at each Cr and three mirror operations perpendicular to $a$, $b$, $c$-axes), all the 48 Cr-Mn bonds related two-spin tensor $P_{ij}^{\alpha\beta\gamma}$ can be deduced from the original $P_{12}^{\alpha\beta\gamma}$.

$P_{ij}^{\alpha\beta\gamma}$ is a polar tensor. Therefore, the general transformation rule for a polar tensor of order 3 is $T'^{ijk}=a_{il}a_{jm}a_{kn}T^{lmn}$, of which $a_{il}$, $a_{jm}$ and $a_{kn}$ are direction cosines (or transformation operators), $T'^{ijk}$ and $T^{lmn}$ are tensor component in new coordinate system and old coordinate system, respectively.

Then we present the transformations of $P_{12}^{\alpha\beta\gamma}$ under five specific symmetry operations in TMCO lattice:

1) The direction cosines of 3-fold rotation along [1,1,1] can be expressed as:

$$\mathbf{3}_{[111]}=\begin{pmatrix} 0 & 1 & 0 \\ 0 & 0 & 1 \\ 1 & 0 & 0 \end{pmatrix} \quad \text{(S6)}$$

Then the transformed $P'^{\alpha\beta\gamma}_{ij}$ can be expressed as (if bond $\langle 1, 2 \rangle$ will be transformed to bond $\langle i, j \rangle$):

$$P'^{\alpha\beta\gamma}_{ij} = \begin{pmatrix} \{P^{c,c,c}, P^{c,c,a}, P^{c,c,b}\} & \{P^{c,a,c}, P^{c,a,a}, P^{c,a,b}\} & \{P^{c,b,c}, P^{c,b,a}, P^{c,b,b}\} \\ \{P^{a,c,c}, P^{a,c,a}, P^{a,c,b}\} & \{P^{a,a,c}, P^{a,a,a}, P^{a,a,b}\} & \{P^{a,b,c}, P^{a,b,a}, P^{a,b,b}\} \\ \{P^{b,c,c}, P^{b,c,a}, P^{b,c,b}\} & \{P^{b,a,c}, P^{b,a,a}, P^{b,a,b}\} & \{P^{b,b,c}, P^{b,b,a}, P^{b,b,b}\} \end{pmatrix} \quad \text{(S7)}$$

2) The direction cosines of $\bar{\mathbf{1}}$ can be expressed as:

$$\bar{\mathbf{1}}=\begin{pmatrix} -1 & 0 & 0 \\ 0 & -1 & 0 \\ 0 & 0 & -1 \end{pmatrix} \quad \text{(S8)}$$

Then the transformed $P'^{\alpha\beta\gamma}_{kl}$ can be expressed as (if bond ⟨*1, 2*⟩ will be transformed

to bond ⟨*k, l*⟩):

$$P'^{\alpha\beta\gamma}_{kl} = \begin{pmatrix} \{-P^{a,a,a}, -P^{a,a,b}, -P^{a,a,c}\} & \{-P^{a,b,a}, -P^{a,b,b}, -P^{a,b,c}\} & \{-P^{a,c,a}, -P^{a,c,b}, -P^{a,c,c}\} \\ \{-P^{b,a,a}, -P^{b,a,b}, -P^{b,a,c}\} & \{-P^{b,b,a}, -P^{b,b,b}, -P^{b,b,c}\} & \{-P^{b,c,a}, -P^{b,c,b}, -P^{b,c,c}\} \\ \{-P^{c,a,a}, -P^{c,a,b}, -P^{c,a,c}\} & \{-P^{c,b,a}, -P^{c,b,b}, -P^{c,b,c}\} & \{-P^{c,c,a}, -P^{c,c,b}, -P^{c,c,c}\} \end{pmatrix} \qquad \text{(S9)}$$

3)  The direction cosines of mirrors perpendicular to *a*-axis, *b*-axis and *c*-axis can

   be expressed as:

$$\boldsymbol{m_a} = \begin{pmatrix} -1 & 0 & 0 \\ 0 & 1 & 0 \\ 0 & 0 & 1 \end{pmatrix} \boldsymbol{m_b} = \begin{pmatrix} 1 & 0 & 0 \\ 0 & -1 & 0 \\ 0 & 0 & 1 \end{pmatrix} \boldsymbol{m_c} = \begin{pmatrix} 1 & 0 & 0 \\ 0 & 1 & 0 \\ 0 & 0 & -1 \end{pmatrix}, \text{respectively (S10)}$$

Then the transformed $P'^{\alpha\beta\gamma}_{mn}$, $P'^{\alpha\beta\gamma}_{op}$, $P'^{\alpha\beta\gamma}_{qr}$ can be expressed as (if bond ⟨*1, 2*⟩ will

be transformed to bond ⟨*m, n*⟩,⟨*o, p*⟩,⟨*q, r*⟩, respectively):

$$P'^{\alpha\beta\gamma}_{mn} = \begin{pmatrix} \{-P^{a,a,a}, P^{a,a,b}, P^{a,a,c}\} & \{P^{a,b,a}, -P^{a,b,b}, -P^{a,b,c}\} & \{P^{a,c,a}, -P^{a,c,b}, -P^{a,c,c}\} \\ \{P^{b,a,a}, -P^{b,a,b}, -P^{b,a,c}\} & \{-P^{b,b,a}, P^{b,b,b}, P^{b,b,c}\} & \{-P^{b,c,a}, P^{b,c,b}, P^{b,c,c}\} \\ \{P^{c,a,a}, -P^{c,a,b}, -P^{c,a,c}\} & \{-P^{c,b,a}, P^{c,b,b}, P^{c,b,c}\} & \{-P^{c,c,a}, P^{c,c,b}, P^{c,c,c}\} \end{pmatrix} \qquad \text{(S11)}$$

$$P'^{\alpha\beta\gamma}_{op} = \begin{pmatrix} \{P^{a,a,a}, -P^{a,a,b}, P^{a,a,c}\} & \{-P^{a,b,a}, P^{a,b,b}, -P^{a,b,c}\} & \{P^{a,c,a}, -P^{a,c,b}, P^{a,c,c}\} \\ \{-P^{b,a,a}, P^{b,a,b}, -P^{b,a,c}\} & \{P^{b,b,a}, -P^{b,b,b}, P^{b,b,c}\} & \{-P^{b,c,a}, P^{b,c,b}, -P^{b,c,c}\} \\ \{P^{c,a,a}, -P^{c,a,b}, P^{c,a,c}\} & \{-P^{c,b,a}, P^{c,b,b}, -P^{c,b,c}\} & \{P^{c,c,a}, -P^{c,c,b}, P^{c,c,c}\} \end{pmatrix} \qquad \text{(S12)}$$

$$P'^{\alpha\beta\gamma}_{qr} = \begin{pmatrix} \{P^{a,a,a}, P^{a,a,b}, -P^{a,a,c}\} & \{P^{a,b,a}, P^{a,b,b}, -P^{a,b,c}\} & \{-P^{a,c,a}, -P^{a,c,b}, P^{a,c,c}\} \\ \{P^{b,a,a}, P^{b,a,b}, -P^{b,a,c}\} & \{P^{b,b,a}, P^{b,b,b}, -P^{b,b,c}\} & \{-P^{b,c,a}, -P^{b,c,b}, P^{b,c,c}\} \\ \{-P^{c,a,a}, -P^{c,a,b}, P^{c,a,c}\} & \{-P^{c,b,a}, -P^{c,b,b}, P^{c,b,c}\} & \{P^{c,c,a}, P^{c,c,b}, -P^{c,c,c}\} \end{pmatrix} \qquad \text{(S13)}$$

All other all Cr-Mn bonds related two-spin tensor $P^{\alpha\beta\gamma}_{ij}$ can be deduced by combing these above symmetry operations.

## 3. Polarization calculation

In these cubic $A$-site ordered perovskite systems with collinear $G$-type antiferromagnetic (AFM) alignments, there are two kinds of AFM configurations connected by space inversion operation at the center of unit cell, as schematically shown in Fig. 3a and Supplementary Fig. 5a.[3] In Supplementary Fig. 5a, the Néel vectors $S_{Mn}$ and $S_{Cr}$ are always antiparallel with each other. The corresponding $P$ vector is opposite between the two AFM configurations when all the spins are aligned along the same diagonal direction. From the symmetry analysis, when magnetic structure points to any diagonal direction, a ferroelectric polarization $P$ along the same diagonal axis will be induced in both spin configurations. When $S_{Mn}$ and $S_{Cr}$ coherently go over all eight equivalent directions, there are eight possible orientations of $P$ vector along $[\pm 1, \pm 1, \pm 1]$.

By assigning AFM configuration of $S_{Cr}$ and $S_{Mn}$ at each site, we can calculate the total polarization $\vec{P}(S_{Mn}, S_{Cr})$ of one magnetic unit cell via

$$P^\gamma(S_{Mn},\ S_{Cr}) = \sum_{<i,j>} P_{ij}^{\alpha\beta\gamma} S_{Mn\alpha} S_{Cr\beta},\ (\alpha,\ \beta = a,\ b,\ c), \qquad (S14)$$

First we assume Néel vector $S_{Cr} = (1/\sqrt{3})\mu_{Cr}[\pm 1, \pm 1, \pm 1]$ and $S_{Mn} = (1/\sqrt{3})\mu_{Mn}[\pm 1, \pm 1, \pm 1]$ along $[\pm 1, \pm 1, \pm 1]$ directions at zero magnetic field. Substituting $S_{Cr}$ and $S_{Mn}$ and Eqs. S7, S9, S11 to S13 into Eq. S14 and summing over one magnetic unit cell of AFM phase, the $\vec{P}(S_{Mn},\ S_{Cr})$ can calculated as:

$$\vec{P}(S_{Mn},\ S_{Cr}) \propto (16/3)\ \mu_{Mn}\mu_{Cr}(P^{a,b,c} + P^{b,c,a} + P^{c,a,b} + P^{a,c,b} + P^{b,a,c} + P^{c,b,a})\ [\pm 1, \pm 1, \pm 1] \quad (S15)$$

From the expression we know that only six components in $P_{12}^{\alpha\beta\gamma}$ contribute to net polarization when $S_{Mn}$// $S_{Cr}$// $[\pm 1, \pm 1, \pm 1]$. For simplicity, we let

$$\Omega = P^{a,b,c} + P^{b,c,a} + P^{c,a,b} + P^{a,c,b} + P^{b,a,c} + P^{c,b,a} \qquad (S16)$$

leading to a simplified Eq. S15 as:

$$\vec{P}(S_{Mn}, S_{Cr}) \propto (16/3)\, \mu_{Mn}\mu_{Cr}\Omega[\pm 1, \pm 1, \pm 1] \qquad (S17)$$

As for the magnetic structure shown in Fig. 3a is concerned, $[\pm 1, \pm 1, \pm 1]$ comprises only combinations with even numbers of the minus sign. Therefore, there are only four among eight diagonal directions are allowed for the polarization vectors. These allowed $P$ vectors form a regular tetrahedron (Fig. 3b). For the spin configuration of Supplementary Fig. 5a, spin-induced $P$ vector can only occupy the rest four directions forming the other tetrahedron (Supplementary Fig. 5b). Along any diagonal directions, the two spin configurations generate opposite $P$ vectors.

To further reveal the spin-polarization relationship when $S_{Cr}$ points to an arbitrary orientation in a 3D space, we assumed $S_{Cr} = (\mu_{Cr}, \theta, \varphi)$ and $S_{Mn} = (\mu_{Mn}, \theta, \varphi)$ (Fig. 3a and c) in a spherical coordinate with $\theta = (0, \pi)$ and $\varphi = (-\pi, \pi)$. By substituting $S_{Cr}$ and $S_{Mn}$ into Eq. S14 and summing over one magnetic unit cell of the AFM phase, the $\vec{P}(S_{Mn}, S_{Cr})$ can be converted into a parameter equation $\vec{P}(\theta, \varphi)$ as follows:

$$\vec{P} = \beta\mu_{Cr}\mu_{Mn}\Omega[\sin(2\theta)\sin(\varphi),\ \sin(2\theta)\cos(\varphi),\ \sin^2\theta\sin(2\varphi)], \qquad (S18)$$

where the coefficient $\beta$ depends on the temperature and magnetic field. The analytical equation of Eq. S18 can be written in a form similar to Eq. 1. For spin configuration shown in Supplementary Fig. 5a, the Eq. S18 becomes:

$$P_a^2 P_b^2 + P_a^2 P_c^2 + P_b^2 P_c^2 + 2\beta_{\mu_{Cr}\mu_{Mn}}\Omega P_a P_b P_c = 0 \qquad (S19)$$

The obtained Eq. S19 and Eq. 3 are fully consistent with Eq. 1 for $\delta = \pm 2\beta\mu_{Cr}\mu_{Mn}\Omega$, respectively. Form Eq. S19, the spin-induced $P$ on the Roman surface is shown in

Supplementary Fig. 5c.

## 4. The Monte Carlo simulations under a small rotating *B*

To describe the external fields *B* dependent spin configurations, we write the following Hamiltonian of a local-moment Heisenberg model by including the AFM interlayer exchange interaction coupling $J<0$, cubic magnetic anisotropy constants $K_1$, $K_2$ and Zeeman energy, the total energy $H$ is given by:

$$H = J\boldsymbol{S}_1 \cdot \boldsymbol{S}_2 + K_1 \sum_i (S_{i,x}^2 S_{i,y}^2 + S_{i,y}^2 S_{i,z}^2 + S_{i,z}^2 S_{i,x}^2) + K_2 \sum_i (S_{i,x}^2 S_{i,y}^2 S_{i,z}^2) - g\mu_B \boldsymbol{B} \cdot \sum_i \boldsymbol{S}_i \quad (S20)$$

where, for the cubic lattice, we omit the higher order anisotropy terms and $i$=1,2 for two antiparallelly aligned Cr spins. Given the easy axis to be [1,1,1], we could determine the relations between parameters:

$$K_2 < -9K_1 \text{ and } K_1 > 0, \text{ or, } 4K_2 < -9K_1 \text{ and } K_1 < 0 \quad (S21)$$

By selecting tentative parameters satisfying the above conditions, $K_1 = -4$, $K_2 = -2$, $J = 20$, $g\mu_B \boldsymbol{B}/S = 8$, Monte Carlo simulations are run with small field rotating in the *ab*-plane as an example, using 30000 Monte Carlo steps per angle to equilibrate the structure and an additional 30000 steps to calculate the average magnetic properties. To be consistent with experimental data, we begin the simulations with $T = 0.1$ K and rotate the angle in equal steps, using the final state from the previous field as the initial state for the current field.

Under small magnetic field, the AFM configuration will cant slightly to induce a ferromagnetic component which is very small comparing to the saturation value while its AFM axis will also deviate from diagonal directions. The dominating change of *P* will mainly come from the deviation of AFM axis (Neel vector $S_{cr}$). As shown in Supplementary Fig. 7, the antiferromagnetic Neel vector $S_{cr} \propto (S_1 - S_2)/2$ is tuned in double period under $B$. The simulated data are fitted with sinusoidal functions in the forms:

$$S_{Cr}^x = 0.54 + 0.03\sin\left[2(\Phi - 48°)\right] \tag{S22}$$

$$S_{Cr}^y = 0.54 + 0.03\sin\left[2(\Phi - 131°)\right] \tag{S23}$$

$$S_{Cr}^z = 0.58 + 0.01\sin\left[2(\Phi + 4°)\right] \tag{S24}$$

where $\Phi$ is the angle of the external field in the $ab$ plane. Then, the corresponding polarizations on the Roman surface have the following form:

$$P_x = 0.63 + 0.04\sin\left[2(\Phi + 40°)\right] \tag{S25}$$

$$P_y = 0.63 - 0.04\sin\left[2(\Phi + 51°)\right] \tag{S26}$$

$$P_z = 0.58 - 0.01\sin\left[2(\Phi + 5°)\right] \tag{S27}$$

Clearly, when the external field rotates 360°, the $S_{Cr}$ and $P$ draw a closed loop twice on the sphere and Roman surface, as shown in Supplementary Fig. 6a and b, respectively. When the small $B$ is rotated along other planes, the situation would be similar.

**Supplementary Table 1 | Typical two-dimensional topological manifolds (surface), their topological invariants, and their realization in physics**. Realization of Roman surface in terms of multiferroicity and magnetoelectricity in TMCO and LaMn$_3$Cr$_4$O$_{12}$ (LMCO).

| 2D topological manifolds | Image | Orientability | Genus | Physical entity | Space |
|---|---|---|---|---|---|
| Sphere $\boldsymbol{S^2}$ | 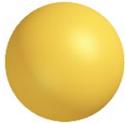 | Orientable | 0 | Skyrmion[5,6], polar skyrmion[7], polar meron[8] | Real |
| Torus $\boldsymbol{T^2}$ | 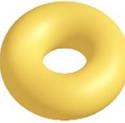 | | 1 | 1st Brillouin zone | Reciprocal |
| Möbius strip $\boldsymbol{M}$ | 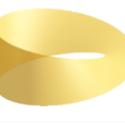 | Non-orientable | 1 | Nanocrystal/DNA band[9,10], Möbius aromaticity in cycloalkene[11] | Real |
| Klein bottle $\boldsymbol{K}$ | 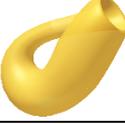 | | 2 | Energy landscape of cyclo-octane[12] | Phase |
| Roman surface $\boldsymbol{RP^2}$ | 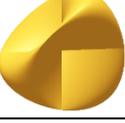 | | 1 | **Multiferroicity/ Magnetoelectricity (This Work)** | Phase |

**Supplementary Table 2 | Structure parameters of TMCO obtained from the Rietveld refinement at room temperature.** The atomic position: Tb 2$a$ (0, 0, 0), Mn 6$b$ (0, 0.5, 0.5), Cr 8$c$ (0.25, 0.25, 0.25), O 24$g$ ($x$, $y$, 0). The bond valence sum (BVS) values ($V_i$) were calculating using the formula $V_i = \sum_j S_{ij}$ and $S_{ij} = \exp[(r_0 - r_{ij})/0.37]$. The parameters of $r_0(\text{Mn}^{3+}) = 1.732$ Å and $r_0(\text{Cr}^{3+}) = 1.708$ Å were used in calculations.

| Parameters | TbMn$_3$Cr$_4$O$_{12}$ |
|---|---|
| $a$ (Å) | 7.35288(1) |
| $V$ (Å$^3$) | 397.532(2) |
| O$_x$ | 0.3062(2) |
| O$_y$ | 0.1763(2) |
| $U_{\text{iso}}$(Tb) (100 × Å$^2$) | 1.13(4) |
| $U_{\text{iso}}$(Mn) (100 × Å$^2$) | 0.66(6) |
| $U_{\text{iso}}$(Cr) (100 × Å$^2$) | 0.26(5) |
| $U_{\text{iso}}$(O) (100 × Å$^2$) | 1.03(6) |
| Mn-O( × 4) (Å) | 1.913(2) |
| Cr-O( × 6) (Å) | 1.965(9) |
| BVS(Mn) | 2.72 |
| BVS(Cr) | 2.99 |
| $R_{\text{wp}}$ (%) | 2.41 |
| $R_{\text{p}}$ (%) | 1.58 |

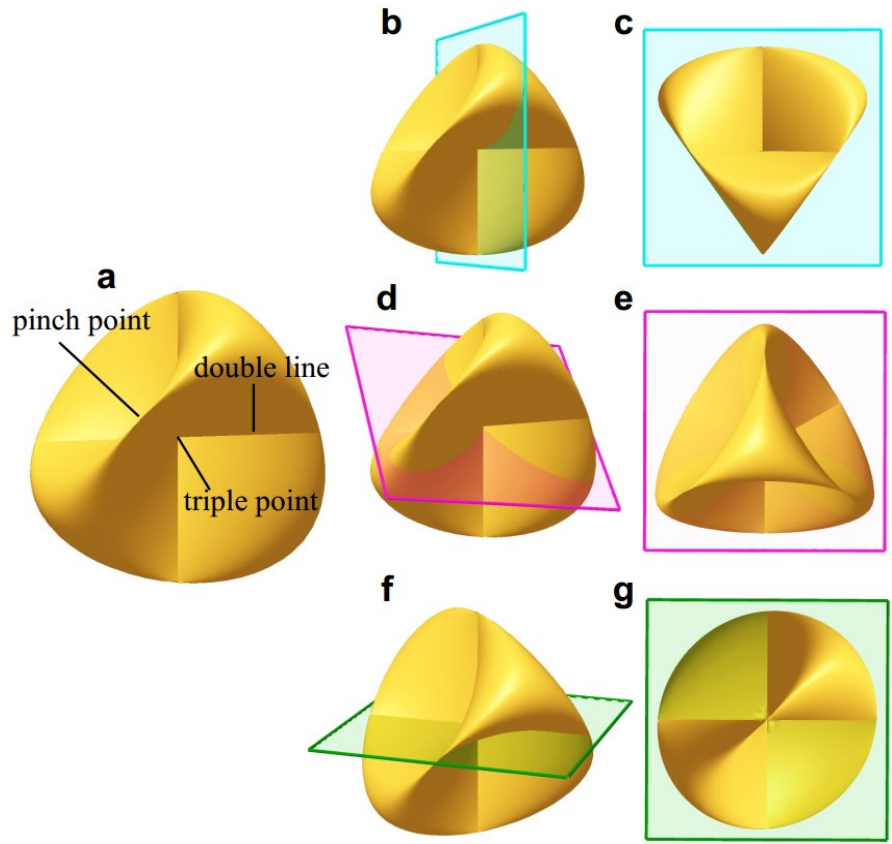

**Supplementary Fig. 1 | Schematic diagram of Roman surface. a**, The diagram of Roman surface with one triple point, six pinch points and three double lines. **b**, **d** and **f**, The different cross sections on Roman surface. Both cross sections in **b** and **d** pass through a double line and two extreme points on the surface. Another cross section in **f** passes through two double lines on the surface. **c**, **e** and **g**, The perspectives perpendicular to cross sections showing in **b**, **d** and **f**, respectively.

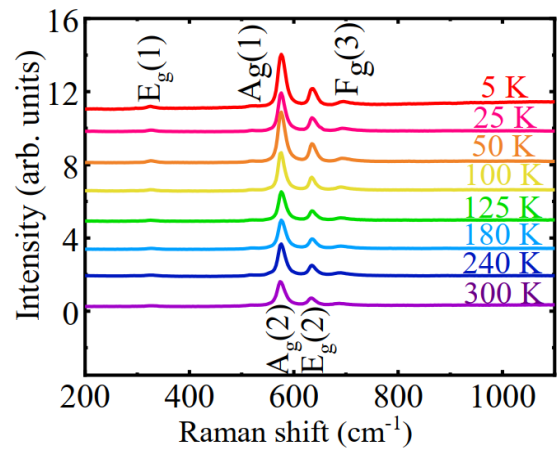

**Supplementary Fig. 2 | Raman spectra measured for TMCO.** Raman spectra measured at different temperatures where different Raman modes are also indicated. No structural phase transition is found to occur with temperature down to 5 K.

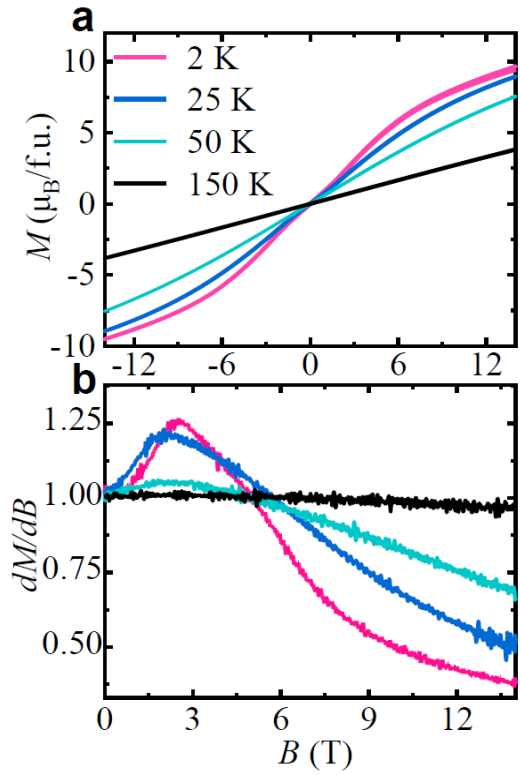

**Supplementary Fig. 3 | Magnetic field dependent magnetization in TMCO. a,** Isothermal magnetization curves measured at selected temperatures. **b,** Magnetic field dependence of the derivation of magnetization. The data are normalized by dividing the value obtained at 0 T. A metamagnetic transition is observed around 2.5 T below $T_{\mathrm{N1}} \approx$ 136 K.

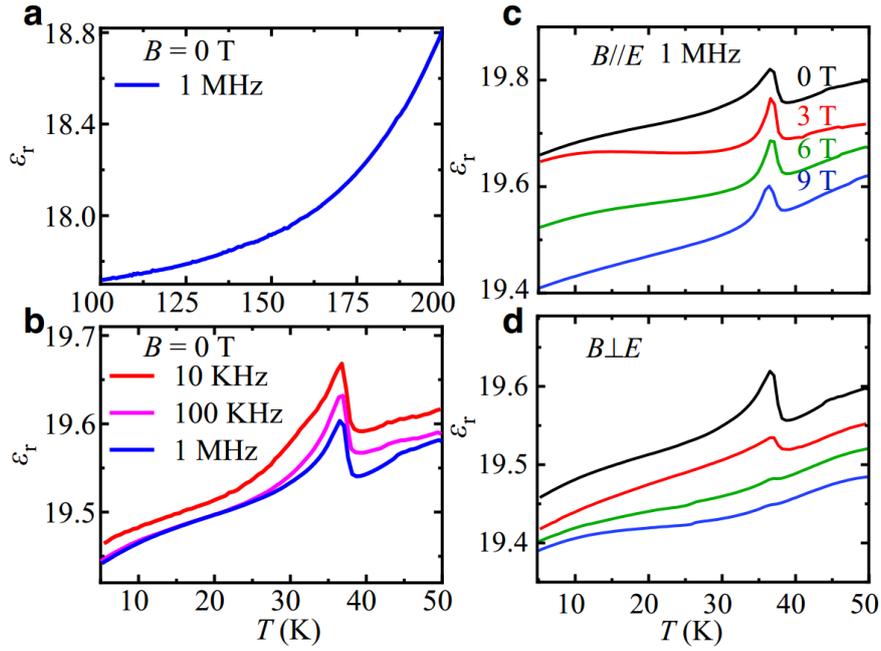

**Supplementary Fig. 4 | Temperature dependence of dielectric constant $\varepsilon_r$ for TMCO. a**, Temperature dependent $\varepsilon_r$ from 100 K to 200 K measured at 1 MHz under zero magnetic field, and the applied electric field $E$ is about 11.7 kV/m. No dielectric anomaly is found to occur at $T_{N1}$. **b**, Temperature dependent $\varepsilon_r$ from 5 K to 50 K measured at different frequencies under zero magnetic field. The frequency independent dielectric peak occurring at $T_{N2}$ suggests a spin-induced ferroelectricity. **c, d**, Temperature dependence of $\varepsilon_r$ measured at 1 MHz under different magnetic fields $B$ applied in parallel and perpendicular to the electric field $E$, respectively. The data are shifted for clarity.

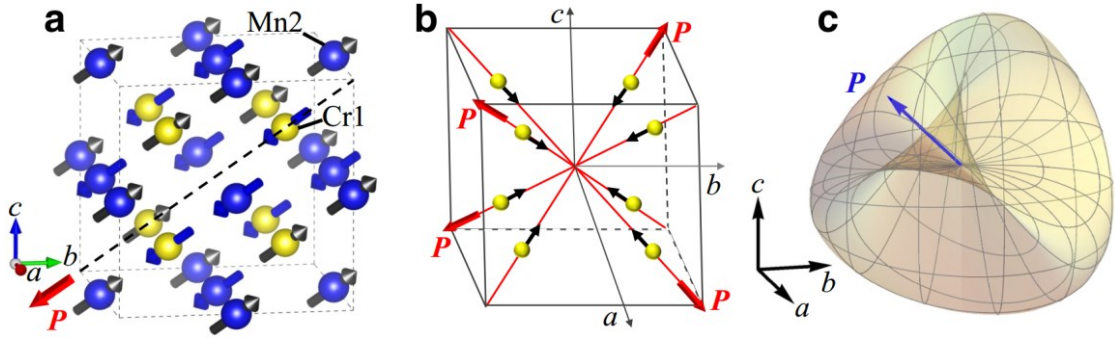

**Supplementary Fig. 5 | Schematic diagram for the other magnetic and ferroelectric structures in $A$Mn₃Cr₄O₁₂. a**, Spin configuration for the other magnetic structure in which Néel vector $S_{Cr}$ is always antiparallel to $S_{Mn}$ while they are parallel in Fig. 3a. This magnetic configuration is linked with that shown in Fig. 3a by the space inversion operation at the center and induces polarization with opposite direction. **b**, Distribution of $P$ vectors along four out of eight diagonal directions with the spin configuration in Supplementary. Fig. 5a. **c**, The complete set of allowed spin-induced $P$ vector in three dimensions by the magnetic structure in the Supplementary Fig. 5a.

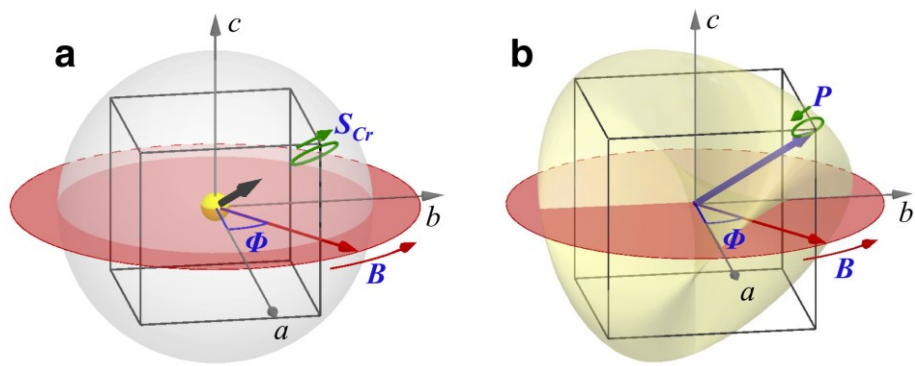

**Supplementary Fig. 6 | Fields dependent loops on sphere of Néel vectors and the induced polarization trajectory on Roman surface. a**, Néel vector rotation mode (green loop) effected by the rotating fields *B*. **b**, Trajectory of *P* in green induced by loops in **a**. As the applied magnetic field rotates 360°, the $S_{Cr}$ and *P* draw a closed loop twice on the sphere and Roman surface, respectively.

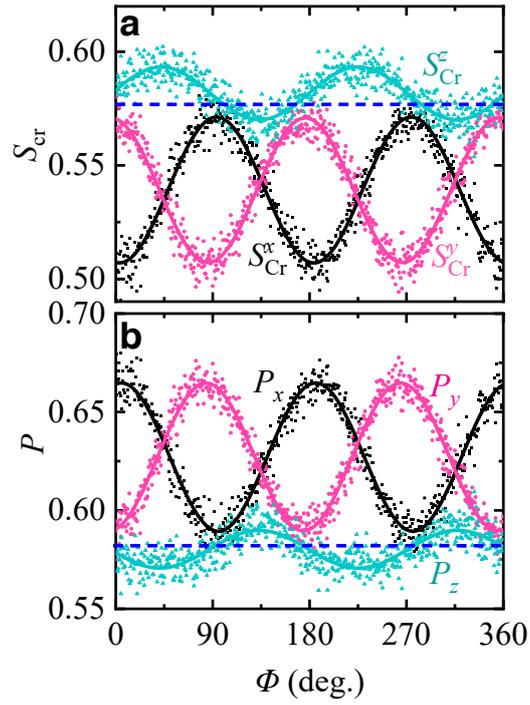

**Supplementary Fig. 7 | Angle $\Phi$ dependent (a) $S_{cr}$ and (b) $P$**. The dots are data from Monte Carlo simulations and the solid lines are fitted sinusoidal curves with double frequency. The Néel vector $S_{Cr}$ and induced $P$ are tuned in double period under $B$.

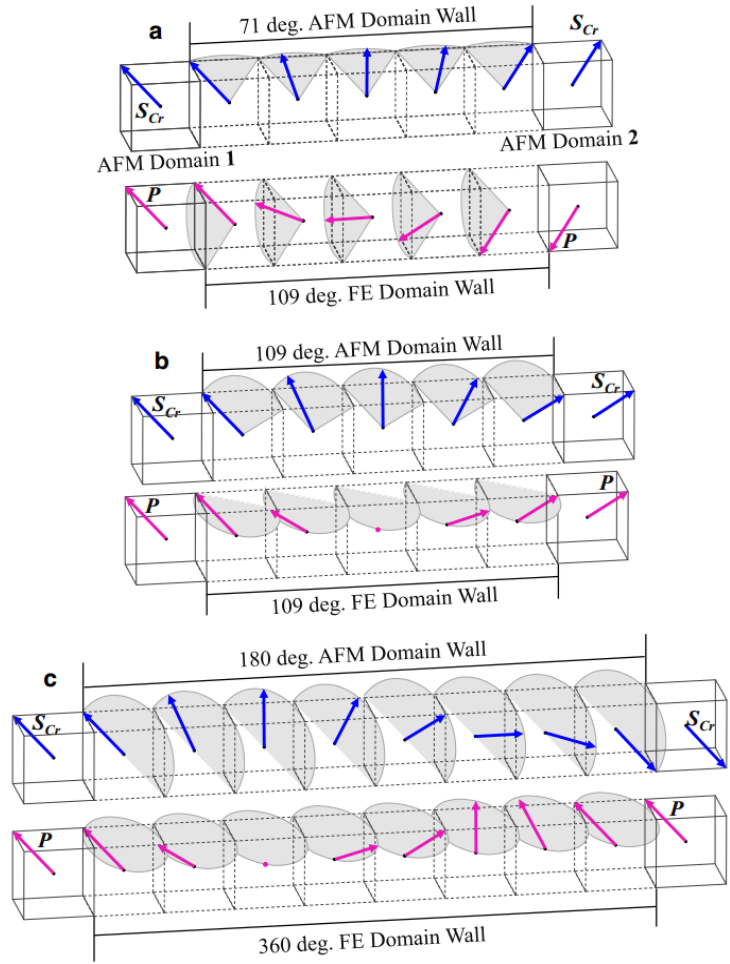

**Supplementary Fig. 8 | Schematic diagram of three types of multiferroic domain walls with $S_{Mn}$ parallel to $S_{Cr}$ in each domain. (a) 71 deg., (b) 109 deg. and (c) 180 deg. AFM domain walls accompanied with corresponding ferroelectric (FE) domain walls by (a) 109 deg., (b) 109 deg. and (c) 360 deg., respectively.** The solid black cells stand for AFM domains and corresponding ferroelectric domains. The dotted black cells are domain walls where shadows indicate rotation planes of Néel vector $S_{Cr}$ and induced $P$. The continuously variation of AFM configuration is represented by the $S_{Cr}$ in blue arrows. The induced $P$ vectors by $S_{Cr}$ inside ferroelectric domain wall are indicated by magenta arrows. In particular, the 180 deg. AFM/360 deg. ferroelectric domain wall would be topologically protected under external electric field.